\documentclass[]{aa}
\usepackage{graphicx}
\usepackage{txfonts}

\begin{document}

   \title{The associated system of HE~2347-4342 \thanks{Based on observations made with the VLT/Kueyen telescope ESO, Paranal, Chile}}

   \author{C. Fechner 
          \and
           R. Baade
          \and
           D. Reimers
          }

  \offprints{C. Fechner}

   \institute{Hamburger Sternwarte, Universit\"at Hamburg,
              Gojenbergsweg 112, D-21029 Hamburg\\
              \email{[cfechner,rbaade,dreimers]@hs.uni-hamburg.de}
             }

   \date{Received 25 September 2003 / Accepted 22 January 2004}

   \abstract{
We present an analysis of the complex associated system of the high-redshift QSO \object{HE~2347-4342}.
Absorption features of \ion{H}{i}, \ion{C}{iii}, \ion{C}{iv}, \ion{N}{v}, and \ion{O}{vi} with up to 16 components occur in the optical spectral range located up to $1500\,\mathrm{km\,s}^{-1}$ redwards from the emission line.
Apparently, \ion{C}{iv} and \ion{N}{v} show the line locking effect. 
A quantivative analysis of the line distribution comparing simulated spectra with randomly distributed doublets reveals, however, no statistical evidence for its physical reality.
Using photoionization calculations to emulate the observed ion column densities we constrain the quasar's spectral energy distribution.
Absorbers in the velocity range of $200 - 600\,\mathrm{km\,s}^{-1}$ can be modelled successfully with a spectral index of $\alpha \sim -3$ at energies higher than $3 - 4\,\mathrm{Ryd}$, which is an energy distribution similar to the QSO continuum suggested by Mathews \& Ferland (\cite{mathewsferland}).
The analysis of a group of high velocity absorbers ($v > 1300\,\mathrm{km\,s}^{-1}$) leads to a harder energy distribution.
The large amount of helium ($\log N_{\mathrm{\ion{He}{ii}}} > 16.3$) associated with these absorbers implies that they are responsible for the observed absence of the proximity effect (Reimers et al. \cite{reimers97}).
Clouds located more distant from the quasar may be shielded from the high energy part of the quasar continuum due to optically thick absorption shortward of $228\,\mathrm{\AA}$ by the high velocity absorbers. 
A group of absorbers with $900 < v < 1200\,\mathrm{km\,s}^{-1}$, in particular a cloud at $1033\,\mathrm{km\,s}^{-1}$, which has the most reliable column density measurements, can be modelled neither with  photoionzation nor under the assumption of collisionally ionized gas.
Possible explanations are a multiphase medium with a mixture of photo and collisionally ionized gas and/or gas in non-equilibrium.
   \keywords{
            quasars: absorption lines --
                quasars: individual: HE~2347-4342 --
		cosmology: observations
               }
   }

   \maketitle
%

\section{Introduction}

Absorption complexes with $z_{\mathrm{abs}} \sim z_{\mathrm{em}}$ are apparently associated with the QSO. 
These so-called associated absorption systems are characterized by highly ionized absorption lines and are generally defined by the criterion that the velocity relative to the quasar is less than $5000\,\mathrm{km\,s}^{-1}$ (Weymann et al. 1979, Foltz et al. 1986, 1988). 
Obviously, the absorbing material belongs to the inner region of active galaxies. 
Studies of associated systems have potentially important implications for galaxy formation and evolution. 
The gas dynamics and velocity fields of the flows in the central region are still a matter of debate.
The metalicities of these systems are solar to several times solar (Wampler et al. 1993). 
High metalicities would agree with predictions of galactic chemical evolution (Hamann \& Ferland 1993).

In contrast to normal intergalactic absorption line systems, which are formed in intervening gas clouds at distances corresponding to their cosmological redshifts, associated systems typically show strong high ionization lines from \ion{O}{vi} and in particular \ion{N}{v} which are rarely detected in intergalactic absorption line systems.
Obviously, the origin of the high ionization is the hard, nonthermal EUV radiation of the parent QSO.
Therefore, column density ratios of \ion{C}{iii}, \ion{C}{iv}, \ion{Si}{iv}, \ion{O}{vi}, and \ion{N}{v} combined with photoionization calculation should -- at least in principle -- allow to estimate the EUV energy distribution of the QSO which is otherwise largely unobservable.

The high-redshift quasar HE~2347-4342 ($z_{\mathrm{em}} = 2.885, V = 16.1$) was
discovered in the course of the Hamburg/ESO Survey (HES; Reimers \& Wisotzki \cite{reimerswisotzki}). 
A strong associated system with absorption components up to $1500\,\mathrm{km\,s}^{-1}$ redwards from the QSO emission line redshift is observed in highly ionized metal lines. 
In this paper we present a detailed analysis of the absorption characteristics and the inferences on the spectral energy distribution of the QSO.

In absorbing gas clouds close to the QSO, radiation pressure due to absorption by strong resonance lines and therefore line locking can play a role, and this appears to happen in the complex associated system of HE~2347-4342.
Line locking occurs if the velocity separation between two absorbing clouds is equal to the velocity separation of a doublet splitting (Scargle \cite{scargle}; Braun \& Milgrom \cite{braunmilgrom}).
Aside from the incidence in stars this effect has been observed mainly in BAL--QSOs (Foltz et al. \cite{foltzetal}; Vilkoviskij \& Irwin \cite{vilkoviskij}).
However, it also occurs in other AGNs with associated systems (e.g. Srianand \cite{srianand}; Srianand et al. \cite{srianandetal}).
With our analysis of the \ion{C}{iv} and \ion{N}{v} complex associated system of HE~2347-4342 we shall perform a statistical test in order to clarify whether the apparent line locking has a physical basis or the line coincidences are within the expectation in a random distribution of a rich associated system with up to 16 components.

The strong associated system of HE~2347-4342 also seems to be responsible for the observed absence of a \ion{He}{ii} proximity effect in this QSO (Reimers et al. \cite{reimers97}) which would imply that at least one of the components of the associated system is optically thick to radiation shortward of $228\,\mathrm{\AA}$.
If the corresponding clouds were closer to the QSO than other components, the radiation shortward of $228\,\mathrm{\AA}$ would be invisible for the more distant absorbers.
This could be observable and our analysis attempts to reveal whether some subcomponents ``see'' a softer ionizing radiation field than others.


\section{Observations}

HE~2347-4342 was observed with the UV-Visual Echelle Spectrograph (UVES) 
at the second VLT Unit Telescope (Kueyen) during four nights in October and 
November 2000.
Twelve individual exposures with integration times of 3600\,s were made using 
the standard settings for the central wavelengths of 3460/4370\,\AA\, in the 
blue and 5800/8600\,\AA\, in the red.
The slit width was 1 arcsec leading to a spectral resolution of about 
40\,000.

The data reduction was performed at Quality Control Garching using the 
UVES pipeline Data Reduction Software (Ballester et al. \cite{ballester}).  
Finally, the vacuum-barycentric corrected spectra were co-added resulting in a signal-to-noise ratio of about 100.

Additional data are available in the wavelength range $1600 - 3300$\,\AA\, obtained with the Faint Object Spectrograph (FOS) onboard of Hubble Space Telescope (HST) in high resolution mode ($R \sim 1300$) on June 7, 1996.
The wavelength range $1150 - 1450$\,\AA\, is covered by observations with the Goddard High Resolution Spectrograph (GHRS) also onboard of HST on June 11 and 14, 1996 with a resolution of 2000 and a total exposure time of 21\,100\,s. 
Because of the poor signal-to-noise ratio we do not use the FOS and GHRS data for a quantitative analysis.

\section{The associated system}

\begin{figure}
   \centering
   \resizebox{\hsize}{!}{\includegraphics[bb=70 65 480 775,clip=]{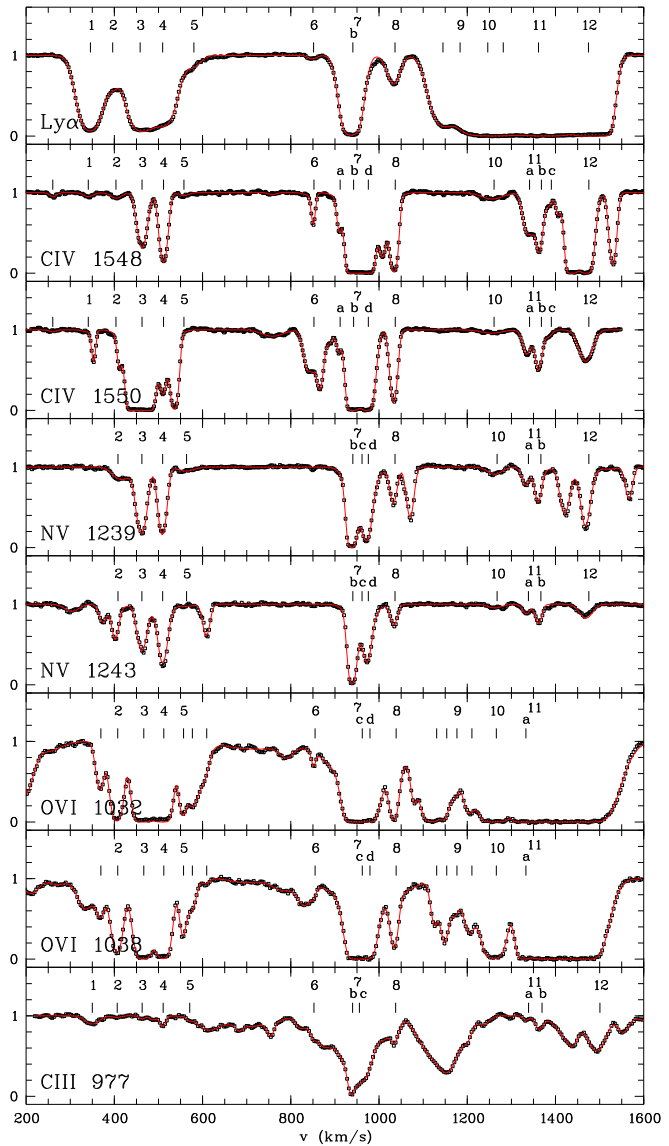}}
      \caption{Absorption line profiles arising in the associated system of HE~2347-4342. The velocity $v = 0 \,\mathrm{km\,s}^{-1}$ corresponds to the QSO redshift $z = 2.885$. The numbers indicate the absorption features included in the analysis of the spectral energy distribution (see Table \ref{columndensities} in Sect. \ref{sed_analysis}).
              }
         \label{asv}
   \end{figure}

As Reimers et al. (\cite{reimers97}) noted, HE~2347-4342 shows no proximity effect, but a strong associated system. 
Absorption lines of \ion{H}{i}, \ion{C}{iv}, \ion{C}{iii}, \ion{N}{v}, and  \ion{O}{vi} observed in the UVES portion of the spectrum, \ion{O}{v}, \ion{Ne}{vii}, and \ion{Ar}{vii} observed with FOS and even \ion{He}{ii} are clearly visible covering a redshift range from 2.877 -- 2.904.
The components are spread up to 1500 $\mathrm{km\, s}^{-1}$ redwards from the QSO redshift.
Within the associated system we distinguish three groups of absorption features.
One group covers the velocity range $200 < v < 600\,\mathrm{km s}^{-1}$, the others span the intervals $800 < v < 1200\,\mathrm{km s}^{-1}$ and $1200 < v < 1600\,\mathrm{km s}^{-1}$, respectively (see Fig. \ref{asv}).

The QSO redshift of $z = 2.885 \pm 0.005$ was estimated by Reimers et al. (\cite{reimers97}) modelling the \ion{O}{i} emission line.
Efforts to improve this value measuring additional emission lines failed, since we do not observe unblended lines of elements in low ionization stages except \ion{O}{i}.
We find $z_{\mathrm{em}} \sim 2.89$ consistent with the adopted value.

The parameters of the observed absorption lines were estimated with the line fitting program CANDALF developed by R. Baade. 
It performs simultaneously a Doppler profile line fit and continuum normalization.
Compared to an apriori continuum determination the continuum level in spectral regions with many lines can be estimated more reliably.
This is an advantage in modelling absorption lines of hydrogen, \ion{C}{iii}, and \ion{O}{vi} which are located in the Lyman $\alpha$ forest, while the continuum is well defined in spectral regimes where the metal components \ion{C}{iv} and \ion{N}{v} arise.

   \begin{figure}
   \centering
   \resizebox{\hsize}{!}{\includegraphics[bb=55 25 540 775,clip=]{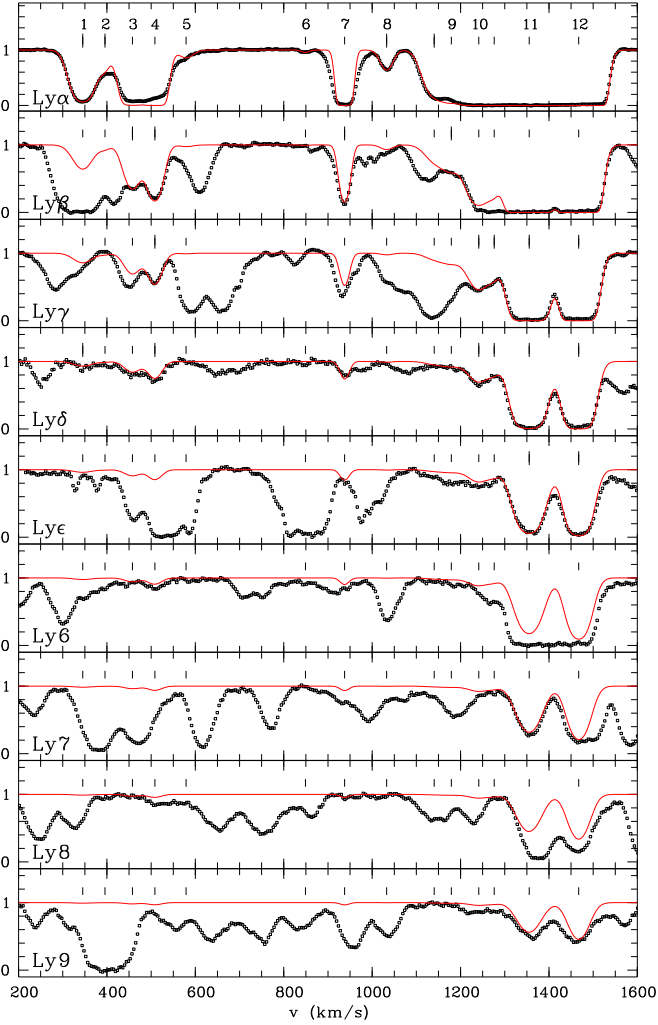}}
      \caption{The \ion{H}{i} Lyman series of the associated system upto Ly9.
The solid line represents the models calculated from the estimated line parameters. The parameter estimation was performed with simultaneous Doppler profile fitting. Absorption features considered in this procedure are indicated by big thickmarks. The small thickmarks indicate the position of the other lines. The numbers refer to Table \ref{clouds} indicating the features included in the SED analysis.
              }
         \label{hicheck}
   \end{figure}

Figure \ref{asv} shows line profiles and fits to the data of the ions visible in the UVES spectrum.
In order to estimate the hydrogen line parameters correctly, we have performed a simultaneous fit considering the Lyman series up to Ly$\epsilon$.
Though higher order lines of the Lyman series are detected, they were not included in the parameter estimation.
Blending with Lyman series lines of a Lyman limit system at $z_{\mathrm{LLS}} \sim 2.735$ and  Ly$\alpha$ forest absorption lines lead to corrupted and thus unusable profiles.  
The hydrogen data and the model are shown in Fig. \ref{hicheck}.

\section{Line Locking}

Line locking is an effect where the velocity separation between two absorbing clouds is equal to the separation (in $\mathrm{km\,s}^{-1}$) between two resonance transitions.
The physical explanation is based on the idea that absorbing clouds are accelerated due to the influence of a central source with resonance line absorption as main acceleration mechanism.
Then line locking of two gas elements is reached by a reduction in the flux that accelerates the more distant gas element because of line absorption produced by the gas closer to the central source.

There are two suggestions how to stabilize the state of line locking, which can be applied to either clouds ejected from the source (Scargle \cite{scargle}) or infalling as well as outflowing material (Braun \& Milgrom \cite{braunmilgrom}) .
The latter would explain the structure of the associated system of HE~2347-4342, if line locking could be confirmed, since its components redwards of the QSO redshift indicate infalling material. 

Considering only absorption features with redshift $z \ge 2.885$, we found 16 doublets of \ion{C}{iv} and 12 of \ion{N}{v} in the spectrum of HE~2347-4342.
Among them there are 6 candidates for line locking in \ion{C}{iv}, and 3 in \ion{N}{v}. 
The complex system of \ion{N}{v} might show another pair of apparently locked doublets, though one is bluewards of the QSO emission redshift at $z = 2.878$.
To restrict these provisional selected candidates to pairs which may be physically locked, we apply a criterion concerning position and width of the involved doublets.
The velocity distance between two doublets should be equal to the velocity splitting $\Delta v_{\mathrm{ion}}$ of the doublet components which is $\sim 499\,\mathrm{km\,s}^{-1}$ for \ion{C}{iv} and $\sim 962\,\mathrm{km\,s}^{-1}$ for \ion{N}{v}.
We formulate an overlap criterion of the form
\begin{equation}\label{criterium}
  \left\vert (v_{2} - v_{1} - \Delta v_{\mathrm{ion}})\right\vert - \frac{1}{2}\left(\mathrm{FWHM}_{1} + \mathrm{FWHM}_{2}\right) < 0\,\mbox{,}
\end{equation}
where $\mathrm{FWHM}_{i} = 2\sqrt{\ln 2} \cdot b_{i}$ is the line width of the participating doublets $i = 1, 2$.
Applying this criterion, the previously selected candidates are confirmed. 

To examine whether the observed line overlaps are only coincidental or statistically evident we introduce the correlation function $\xi(\Delta v)$, which compares the number $n_{\mathrm{obs}}$ of observed overlapping doublets satisfying Eq. (\ref{criterium}) with the expected number $n_{\mathrm{exp}}$  of coincidences in a randomly generated spectrum:
\begin{equation}
  \xi(\Delta v) = \frac{n_{\mathrm{obs}}}{n_{\mathrm{exp}}} - 1 \,\mbox{.}
\end{equation}
We derive $n_{\mathrm{exp}}$ from Monte Carlo simulations.
In accordance with observations, a number of doublets was distributed randomly over a redshift range $2.87 \le z \le 2.91$.
We also simulate a random set of corresponding $b$-parameters matching the observations with $3 \le b \le 30\,\mathrm{km\,s}^{-1}$. 
The number of pairs satisfying our line locking criterion are counted.
This procedure is repeated 10\,000 times for \ion{C}{iv} and \ion{N}{v}. 

   \begin{table}
      \caption[]{Results of the Monte Carlo Simulation. Correlation between the number of observed doubletpairs satisfying Eq. (\ref{criterium}) $n_{\mathrm{obs}}$ and the simulated number of coincidences $n_{\mathrm{exp}}$ with $b$-parameters in the given range. $\xi$ denotes the correlation function and $\sigma_{\xi = 0}$ its $1 \sigma$- Poisson error.}
         \label{montecarlo}
     $$ 
         \begin{array}{l c c c c c}
            \hline
            \noalign{\smallskip}
             \mathrm{Ion} & n_{\mathrm{obs}} & n_{\mathrm{exp}} & \xi & \sigma_{\xi = 0} & b\,(\mathrm{km\,s}^{-1}) \\
            \noalign{\smallskip}
            \hline
            \noalign{\smallskip}
\ion{C}{iv} & 6.0 & 5.1901 & 0.1560 & 0.4913 & 3 \dots 30 \\
\ion{C}{iv} & 6.0 & 3.6030 & 0.6653 & 0.8339 & 3 \dots 20 \\
\ion{N}{v}  & 3.0 & 2.2047 & 0.3607 & 0.8303 & 3 \dots 30 \\
\ion{N}{v}  & 3.0 & 1.8650 & 0.6086 & 1.0408 & 3 \dots 25 \\
\ion{N}{v}  & 3.0 & 2.1380 & 0.4032 & 0.8662 & 7 \dots 25 \\
            \noalign{\smallskip}
            \hline
         \end{array}
     $$ 
   \end{table}

As can be seen from Table \ref{montecarlo}, the number of expected coincidences depends on the line width interval.
The broadest \ion{C}{iv} absorption feature is a weak doublet at $v = (1260.6 \pm 5.7)\,\mathrm{km\,s}^{-1}$ with  $b = (33.1 \pm 12.7)\,\mathrm{km\,s}^{-1}$, which is not among the line locking candidates.
Also the broadest \ion{N}{v} doublet, $b = (30.4 \pm 2.2)\,\mathrm{km\,s}^{-1}$ at $v = (961.3 \pm 1.8)\,\mathrm{km\,s}^{-1}$, which is strongly blended but clearly identified by our profile fitting procedure, shows apparently no line locking effect.
Most of the observed doublets are at least $10$ (\ion{C}{iv}) or $5\,\mathrm{km\,s}^{-1}$ (\ion{N}{v}) narrower.
Therefore we simulate doublets within several ranges of $b$-parameters.
The \ion{N}{v} simulation with $7\,\mathrm{km\,s}^{-1}$ as lower limit is included because the narrowest doublet has a width of $b = (8.2 \pm 0.8)\,\mathrm{km\,s}^{-1}$ while the narrowest \ion{C}{iv} component shows $b = (4.1 \pm 0.9)\,\mathrm{km\,s}^{-1}$ 

We find values of the correlation function only slightly above $0$ in all cases.
Thus, we conclude that there is no statistical evidence for line locking in the associated system of HE~2347-4342.
As expected, the number of simulated conincidences increases if the $b$-parameter range includes broader lines.
But the given intervals do not lead to significantly larger values of $\xi$.
In addition, there are restrictions to the number of observed \ion{C}{iv} coincidences, since saturated lines are involved in 3 out of 6 doublet pairs.
Saturation might cause ambiguous results from Doppler profile fitting, i.e. the column density might be overestimated while the $b$-parameter is too small or vice versa.
Excluding the two saturated \ion{C}{iv} doublets and performing Monte Carlo simulation again yields 2.82 (4.06) simulated coincidences with line widths $3.0 \le b \le 20.0\,(30.0)\,\mathrm{km\,s}^{-1}$ in comparison to 3 observed pairs of doublets.

Smette et al. (\cite{smette}) also noted the apparent state of line locking, which seemed to them surprising. 
We confirm the apparent state implying the line locking effect is present but we found no statistical evidence.
Thus, the observed structure of absorbers is consistent with a stochastic origen and we cannot draw conclusions on the role of radiative acceleration in the kinematics of the absorbing clouds.

\section{Spectral energy distribution}
\label{sed_analysis}

\begin{figure}
   \centering
   \resizebox{\hsize}{!}{\includegraphics[angle=-90]{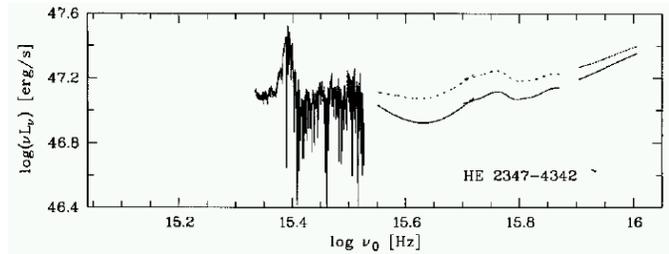}}
      \caption{Observed spectral energy distribution $\log (\nu L_{\nu})\, [\mathrm{erg\,s}^{-1}]$ versus rest frequency of HE~2347-4342. The UV continuum derived for the dereddened spectrum has been corrected for continuum absorption of the Lyman limit system. An additional correction for the cumulative \ion{H}{i} continuum absorption of Ly$\alpha$ clouds with $\log N(\ion{H}{i}) \le 16 \,\mathrm{cm}^{-2}$ is indicated by the dotted line (adopted from Reimers et al. \cite{reimers98}).
              }
         \label{sed}
   \end{figure}

Direct observations of the spectral energy distribution (SED) of QSOs are difficult. 
Especially the position of the so called ``big blue bump'' is hidden by galactic absorption.
To get more information composite spectra are generated by co--adding different spectra in the AGN rest frame (e.g. Zheng et al. \cite{zheng}; Telfer et al. \cite{telfer}).
Even though there are hints that a peak of the energy spectrum of quasars occurs near the Lyman limit (an example is the $z = 0.158$ quasar 3C273, Kriss et al. \cite{krissetal}), individual sources can show substantially different energy distributions in the extreme-UV (e.g. Reimers et al. \cite{reimers98}).  
In the case of HE~2347-4342 Telfer et al. (\cite{telfer}) found an extremly hard EUV continuum (i.e. $\alpha = 0.56$) at rest-frame wavelengths of about $500 - 1200\,\mathrm{\AA}$.
Reimers et al. (\cite{reimers98}) observed the SED of HE~2347-4342 up to a frequency of $\log \nu_{0} = 16$, which corresponds to a wavelength of $300\,\mathrm{\AA}$, in the QSO rest frame.
Part of their figure 2 is shown in our Fig. \ref{sed}.
Obviously, the SED increases up to the highest observed frequencies and the turnover does not occur until a rest frame frequency of $10^{16.0}\,\mathrm{Hz}$, which corresponds to an energy of about $3.0\,\mathrm{Ryd}$.

An associated system provides the opportunity to constrain the SED indirectly.
The task is to model the column densities of observed ions by means of a photoionization code assuming a suitable QSO spectrum that matches the observed part of the SED.
The more different ions are observed the more reliably the energy distribution can be modelled.
Abundance effects can be avoided or at least minimized if several ionization stages of the same element are observable.
The associated system investigated here shows features of carbon in two different ionization stages for at least 6 absorbers.

\subsection{Remarks on individual systems}
\label{remarks}

We concentrate our study on 12 absorbing components detected in the associated system, which are visible at least in two different ions of \ion{H}{i}, \ion{C}{iv}, \ion{N}{v}, and \ion{O}{vi}.
Absorption features of different ions with a velocity separation of less than $10\,\mathrm{km\,s}^{-1}$ were assumed to arise from the same cloud. 
Table \ref{clouds} summarizes the observed ions and their mean velocity.
Apart from absorbers 7 and 11, one feature per ion is observed for every absorbing cloud. 
Number 7 shows saturated lines of \ion{C}{iv} and \ion{O}{vi} and at least two components of \ion{N}{v}.
We add up all column densities in the velocity range $900 < v < 1000 \,\mathrm{km\,s}^{-1}$ to one artificial absorber.
Absorption in the velocity range of $1300 - 1400\,\mathrm{km\,s}^{-1}$ (number 11) consists of at least two components visible in \ion{C}{iv} and \ion{N}{v}. 
The corresponding \ion{H}{i} lines are blended and even in Ly$\epsilon$ one component is still saturated.
For each ion we use the total column density of all components in this velocity range.
For non-detectable ions an upper limit is derived.
Altogether, components of \ion{H}{i}, \ion{C}{iv}, \ion{C}{iii}, \ion{N}{v}, \ion{O}{vi}, and \ion{Si}{iv} are considered in this analysis.
Their measured column densities are listed in Table \ref{columndensities}.
  
   \begin{table}
      \caption[]{Mean absorber position and ions included in the SED analysis. The given velocity is the weighted mean of the position of all visible ions estimated by Doppler profile fitting, or the center of the velocity interval in which all column densities are added (Absorbers number 7 and 11). Ions in parenthesis are detected but not considered in the SED fitting procedure.}
         \label{clouds}
     $$ 
         \begin{array}{c c l }
            \hline
            \noalign{\smallskip}
             \mathrm{Absorber} & v\,(\mathrm{km\,s}^{-1} ) & \mathrm{Ions} \\
            \noalign{\smallskip}
            \hline
            \noalign{\smallskip}
1 &  345.1 \pm  1.3 & \ion{H}{i}, \ion{C}{iv}, \ion{C}{iii} \\
2 &  405.1 \pm  2.0 & \ion{H}{i}, \ion{C}{iv}, \ion{C}{iii}, \ion{N}{v}, \ion{O}{vi} \\
3 &  462.1 \pm  0.7 & \ion{H}{i}, \ion{C}{iv}, \ion{C}{iii}, \ion{N}{v}, \ion{O}{vi}^{\mathrm{a}} \\
4 &  509.7 \pm  0.5 & \ion{H}{i}, \ion{C}{iv}, \ion{C}{iii}, \ion{N}{v}, \ion{O}{vi}^{\mathrm{a}} \\
5 &  560.8 \pm  4.3 & \ion{H}{i}, \ion{C}{iv}, (\ion{C}{iii})^{\mathrm{b}}, \ion{N}{v}, \ion{O}{vi} \\
6 &  850.5 \pm  0.6 & \ion{H}{i}, \ion{C}{iv}, (\ion{C}{iii})^{\mathrm{b}}, \ion{O}{vi} \\
7 &  950.0 \pm 50.0 & \ion{H}{i}, \ion{C}{iv}^{\mathrm{a}}, (\ion{C}{iii}), \ion{N}{v}, \ion{O}{vi}^{\mathrm{a}} \\
8 & 1033.3 \pm  0.4 & \ion{H}{i}, \ion{C}{iv}, \ion{C}{iii}, \ion{N}{v}, \ion{O}{vi} \\
9 & 1173.4 \pm  3.0 & \ion{H}{i}, \ion{O}{vi} \\
10 & 1258.9 \pm  3.0 & \ion{H}{i}, \ion{C}{iv}, \ion{N}{v}, \ion{O}{vi}^{\mathrm{a}} \\
11 & 1350.0 \pm 50.0 & \ion{H}{i}^{\mathrm{a}}, \ion{C}{iv}, \ion{C}{iii}, \ion{N}{v}, (\ion{O}{vi})^{\mathrm{a}} \\
12 & 1467.8 \pm  0.4 & \ion{H}{i}^{\mathrm{a}}, \ion{C}{iv}, (\ion{C}{iii})^{\mathrm{b}}, \ion{N}{v}, (\ion{O}{vi})^{\mathrm{a}} \\
            \noalign{\smallskip}
            \hline
         \end{array}
     $$ 
\begin{list}{}{}
\item[$^{\mathrm{a}}$] saturated
\item[$^{\mathrm{b}}$] blended
\end{list}
   \end{table}
   \begin{table*}
      \caption[]{Measured column densities of the absorbers considered in the SED analysis. Absorbers 7 and 11 consist of several sub-systems (a, b, c, d, see Fig. \ref{asv}), whose column densities were added to one value.}
         \label{columndensities}
     $$ 
         \begin{array}{c c c c c c c }
            \hline
            \noalign{\smallskip}
             \mathrm{Absorber} & \log N_{\mathrm{HI}} & \log N_{\mathrm{\ion{C}{iv}}} &  \log N_{\mathrm{\ion{C}{iii}}} &  \log N_{\mathrm{\ion{N}{v}}} &  \log N_{\mathrm{\ion{O}{vi}}} & \log N_{\mathrm{\ion{Si}{iv}}} \\
            \noalign{\smallskip}
            \hline
            \noalign{\smallskip}
1  & 14.03 \pm 0.01 & 12.25 \pm 0.21 & 12.33 \pm 0.25 &<11.62          &<12.15          & <11.07 \\
2  & 12.92 \pm 0.13 & 12.44 \pm 0.15 & 11.40 \pm 1.60 & 12.90 \pm 0.12 & 14.50 \pm 0.03 & <11.08 \\
3  & 14.33 \pm 0.08 & 13.57 \pm 0.01 & 11.52 \pm 1.21 & 13.93 \pm 0.01 & 15.04 \pm 0.08 & <11.07 \\
4  & 14.49 \pm 0.07 & 13.72 \pm 0.02 & 12.08 \pm 0.46 & 13.87 \pm 0.02 & 14.93 \pm 0.07 & <11.08 \\
5  & 12.64 \pm 0.15 & 12.12 \pm 0.29 &<11.29          & 12.64 \pm 0.22 & 14.06 \pm 0.10 & <11.07 \\
6  & 11.88 \pm 0.28 & 12.87 \pm 0.04 &<11.55          &<11.65          & 12.85 \pm 0.18 & <11.08 \\
7  & 14.34 \pm 0.13 & 15.09 \pm 0.12 &<13.91          & 14.08 \pm 0.07 & 15.34 \pm 0.07 & <11.08 \\
~~\mathrm{a} & ... & 13.05 \pm 0.05 & ... & ...            & ...            & ... \\ 
~~\mathrm{b} & ... & 14.90 \pm 0.18 & ... & 14.53 \pm 0.09 & ...            & ... \\
~~\mathrm{c} & ... & ...            & ... & 14.02 \pm 0.11 & 15.48 \pm 0.14 & ... \\
~~\mathrm{d} & ... & 14.64 \pm 0.11 & ... & 13.83 \pm 0.10 & 14.60 \pm 0.06 & ... \\ 
8  & 13.05 \pm 0.04 & 14.01 \pm 0.02 & 12.70 \pm 0.31 & 13.36 \pm 0.03 & 14.25 \pm 0.04 & <11.08 \\
9  & 14.02 \pm 0.65 &<11.42          &      -         &<11.64          & 13.92 \pm 0.26 & <11.08 \\
10 & 14.77 \pm 0.19 & 12.80 \pm 0.25 &      -         & 12.86 \pm 0.12 & 15.08 \pm 0.05 & <11.07 \\
11 & 15.95 \pm 0.02 & 13.76 \pm 0.03 & 12.41 \pm 0.27 & 13.46 \pm 0.04 &(>14.92)        & <11.07 \\
~~\mathrm{a} & ... & 13.32 \pm 0.04 & 11.65 \pm 0.86 & 13.01 \pm 0.07 & ... & ... \\
~~\mathrm{b} & ... & 13.51 \pm 0.03 & 12.33 \pm 0.27 & 13.27 \pm 0.04 & ... & ... \\
~~\mathrm{c} & ... & 12.64 \pm 0.18 & ...            & ...            & ... & ... \\
12 & 16.02 \pm 0.03 & 13.66 \pm 0.03 &<13.14          & 13.42 \pm 0.04 &(>14.62)        & <11.11 \\
            \noalign{\smallskip}
            \hline
         \end{array}
     $$ 
   \end{table*}

Since our method depends crucially on the column density values and their errors, we add some remarks about the reliability of the derived values and their effects on the estimation of the SED.

Absorber 1: An unblended Ly$\alpha$ line is detected with well-determined column density.
The carbon features are only weak but both \ion{C}{iv} and \ion{C}{iii} appear to be unblended.

Absorber 2: The weak hydrogen feature is blended.
Thus, its column density is not well defined.
Carbon and nitrogen are weak as well but at least unblended in one component. 
The corresponding \ion{O}{vi} doublet starts to be saturated, but nevertheless the $b$-parameter and therefore the column density can still be estimated confidently.

Absorbers 3 \& 4: The Ly$\alpha$ lines are severly blended, but the corresponding Ly$\beta$ features are resolved well.
\ion{O}{vi} suffers from saturation in both components.
Nevertheless, they can be modelled with reasonable line widths -- $(17.7 \pm 1.4)\,\mathrm{km\,s}^{-1}$ for number 3 and $(16.8 \pm 1.6)\,\mathrm{km\,s}^{-1}$ for number 4, so that even the derived column density values are considered to be reliable.
One doublet component of \ion{C}{iv} and \ion{N}{v}, respectively, is unsaturated and unblended.

 Absorber 5: The weak and partly blended \ion{H}{i} absorption feature leads to an uncertain hydrogen column density in this case.
\ion{C}{iv} and \ion{N}{v} show only weak features as well without having line blending problems. 
\ion{C}{iii} is severly blended and cannot be identified unambiguously.
Thus, it is not included in the fit.

Absorber 6: A weak but unblended \ion{H}{i} feature is detected.
Features of \ion{C}{iv} and \ion{O}{vi} can be clearly identified at least in the first doublet component.
\ion{C}{iii} is severly blended and therefore excluded from the fit.

Absorber 7: The hydrogen column density is estimated using the Ly$\beta$ feature since Ly$\alpha$ is saturated.
Both components of \ion{C}{iv} as well as \ion{O}{vi} suffer from saturation, which may lead to uncertainties of the estimated column densities of this combined absorber complex.
\ion{C}{iii} is heavily blended and not included in the fit.
The estimation of the \ion{N}{v} column density is straight forward since at least one component per doublet is unsaturated.
All metal lines show absorption features of at least two components.
For the analysis we use the added column density values. 

Absorber 8: Unblended and unsaturated features are observed in all detectable ions.
Therefore, it is supposed to be an excellent absorber for our attempt to model the radiation field.

Absorber 9: Since the hydrogen absorption feature is blended in Ly$\alpha$ as well as in higher order Lyman lines, its column density is rather uncertain.
The first doublet component of \ion{O}{vi} is blended with Ly$\alpha$ forest lines, but the second is clearly detectable.
Absorption features of carbon and nitrogen are not detected.
Thus, this absorber provides only little information to constrain the SED.

Absorber 10:  Only weak features of \ion{C}{iv} and \ion{N}{v} are detected.
\ion{C}{iii} cannot be isolated and \ion{O}{vi} is saturated in both components.
Also Ly$\alpha$ is saturated and even blended. 
Therefore, the estimation of the hydrogen column density is performed using the unsaturated features of Ly$\gamma$ and Ly$\delta$.

Absorbers 11 \& 12: Strong hydrogen absorption is observed. 
The blend of the two components is resolved for Ly$\gamma$ and higher orders, and in Ly$\epsilon$ the features appear only slightly saturated.
For this reason, we fit the Lyman series from Ly$\gamma$ to Ly$\epsilon$ for the column density determination.
As can be seen from Fig. \ref{hicheck}, the estimated parameters are consistent with the observed line profile up to Ly 9.
The absorption features of \ion{C}{iv} and \ion{N}{v} are unblended in at least one doublet component.
Absorber 11 shows at least two sub-systems in carbon and nitrogen, whose column densities values are added up for analysis.

More crucial is the detection of \ion{O}{vi}, because two heavily saturated absorption troughs are observed at the corresponding wavelengths.
Since there is a metal line system with \ion{C}{iv}, \ion{Si}{iv} and \ion{Si}{iii} observed at $z \sim 2.3132$, at least the first absorption feature contains hydrogen as well.
The corresponding Ly$\beta$ is located in the portion of the spectrum where the flux is extremely reduced due to a Lyman limit system on the line of sight.
Despite the low signal-to-noise ratio an absorption feature at the exspected position is clearly detected.
A simultaneous fit of Ly$\alpha$ and Ly$\beta$ reveals that the features can be described by hydrogen absorption with a column density of $\log N = 15.30 \pm 0.09$ and a Doppler parameter of $b = (75.3 \pm 3.4)\,\mathrm{km\,s}^{-1}$, requiring no additional absorption from metal lines.
Proceeding analogously with the second absorption trough, we find $\log N = 15.14 \pm 0.14$ and $b = (55.6 \pm 3.9)\,\mathrm{km\,s}^{-1}$ for Ly$\alpha$ and Ly$\beta$ at $z \sim 2.3318$.
In this case Ly$\beta$ is located as well in the low flux part of the spectrum, bluewards from the Lyman limit, and is blended with a strong lower redshifted Ly$\alpha$ forest line.

One should keep in mind that both considered \ion{H}{i} lines are heavily saturated and the portion of the spectrum with Ly$\beta$ suffers from a poor signal-to-noise ratio ($S/N \sim 7$).
Thus, the existence of at least one or two \ion{O}{vi} doublets in the Ly$\alpha$ absorption troughs cannot be ruled out completely. 
The column densities derived for the \ion{O}{vi} doublets of absorbers 11 and 12 are based on the assumption that both components are saturated.
We consider the derived values as limits and consequently not include them into the data for the fitting procedure described in Sect. \ref{fitprocedure}.

\subsection{Model assumptions and fitting procedure}
\label{fitprocedure}

Using the photoionization code CLOUDY 94.00 (Ferland \cite{cloudy}) we compute grids of models for the considered column densities.
An absorbing cloud is modelled as a plan-parallel slab of constant density illuminated on one side by the ionizing radiation field.
As input parameters the shape of the QSO continuum, element abundances, total hydrogen density $n_{\mathrm{H}}$ and the ionization parameter $U$ are required.
We adopt a spectral energy distribution parameterized as power law $f_{\nu} \propto \nu^{\alpha}$ with turnover energies $E_{\mathrm{t}}$ between $2.0 - 25.0 \,\mathrm{Ryd}$ and slopes in the range of $0.0 \ge \alpha \ge -9.0$.
Thus, the incident continuum has the form:
\[
  f_{\nu} \propto \left\{ \begin{array}{lll}
    \nu^{-1.0}    & \quad (E \le 2.0 \,\mathrm{Ryd}) \mbox{\,,} \\
    \nu^{+0.5}    & \quad (2.0 \,\mathrm{Ryd} \le E \le E_{\mathrm{t}}) \mbox{\,,} \\
    \nu^{\alpha}  & \quad (E_{\mathrm{t}} \le E \le 10^{3}\,\mathrm{Ryd}) \mbox{\,,}\\
    \nu^{-2.0}    & \quad (10^{3}\,\mathrm{Ryd} \le E) \mbox{\,.}
  \end{array} \right.
\]
The low energy part of the assumed distribution mimics the observed part of the SED shown in Fig. \ref{sed}.
HE~2347-4342 is observed with HST instruments up to 3 Ryd and with the Far Ultraviolet Spectroscopic Explorer (FUSE) up to 3.5 Ryd (Kriss et al. \cite{krissetal2001}) showing no turnover in the obtained data.
Thus, realistic models should lead to a turnover energy of $E_{\mathrm{t}} > 3\,\mathrm{Ryd}$.

The elemental abundances are assumed to be solar and are taken from Grevesse \& Sauval (\cite{grevessesauval}).
Solar metalicities or even higher values seem to be typical for QSO environments (Hamann \& Ferland \cite{hamannferland}). 
The total hydrogen density is set to $10^{-2}\,\mathrm{cm}^{-3}$.
Test calculations with CLOUDY show that the considered column densities weakly depend on this parameter in the range of $-4 \le \log n_{\mathrm{H}} \le 3$.  
The slab geometry as well as the constant density assumption serves as a first step idealization.
The true physical conditions in this multicomponent system are probably more complicated and density fluctuations may be relevant.

The ionization parameter is defined as the ratio of the hydrogen-ionizing photon density to the hydrogen density:
\begin{equation}
  U = \frac{Q(\mathrm{H})}{4 \pi r^{2} n_{\mathrm{H}} c} \mbox{\,,}
\end{equation}
where $Q(\mathrm{H})$ is the number of hydrogen ionizing photons emitted by the central source and $r$ is the distance between source and the illuminated face of the cloud. 
The ionization parameter fixes the radiation flux at the location of the cloud.
It is varied as a model parameter in the range $-4.0 \le \log U \le 1.0$.

We fit the observed column density ratios to our modelgrids. 
In this step turnover and slope of the incident spectrum are still fixed and only the ionization parameter is varied.
The fit is performed as a $\chi^{2}$- minimization.
All observed column density ratios of metal lines to hydrogen contribute to the $\chi^{2}$- function weighted by their errors resulting from the line profile fitting procedure.
Thus, the function to be minimized is
\begin{equation}
  \chi^{2} = \sum_{i=1}^{m} \left(\frac{\log\left(N_{i,\mathrm{obs}}/N_{\mathrm{\ion{H}{i}, obs}}\right) - \log\left(N_{i, \mathrm{mod}}/N_{\mathrm{\ion{H}{i}, mod}} \right)}{\sqrt{(\Delta \log N_{i, \mathrm{obs}})^{2}+(\Delta \log N_{\mathrm{\ion{H}{i}, obs}})^{2}}}\right)^{2}\mbox{,}
\end{equation}
where $m$ is the number of observed metal ions per absorber.
The observed upper limits give an opportunity to check the obtained results a posteriori.
For every absorber we yield optimized ionization parameters and $\chi^{2}$-values for each combination of turnover energy and slope.

Fitted column density ratios are retransformed into column density values, which can be compared directly to the observed quantities.
A comparison between observed limits and modelled column densities is performed to confirm the quality of the fit.
Therefore we compute the fraction $x$ describing the difference between derived
column densities and their limits:
\begin{equation}
  x_{i} =  \frac{\log N_{\mathrm{fit},\,i}}{\log N_{\mathrm{lim},\,i}} - 1 \mbox{.}
\end{equation}
This is done for all observed limits $i$ of every absorber.
The fraction $x$ is negativ if the modelled column densities are smaller than the derived limits.
In the following, the limit consistency (LC) check refers to this equation.

To quantify the quality of the fit we compute a goodness-of-fit parameter $Q (\chi^{2}|\nu)$, the probability that the observed minimum is larger than the value $\chi^{2}$ even if the model is correct.
Here, $\nu$ denotes the number of degrees of freedom, which we set to $n -1$, where $n$ is the number of observed ions per absorber including \ion{H}{i}.
A goodness-of-fit parameter near 1 would confirm the correspondig model as a reasonable description, while $Q \sim 0$ implies that the model is most probably incorrect.

Our models do not consider contributions of other radiation sources like the intergalactic background and local sources (e.g. stars).
It is assumed that other radiation sources are negligible for the associated system.
The modelled SEDs concentrate on the observed energy range and the slope just after the turnover energy.
Contributions of the hard X-ray band are not considered.
The SED of 3C~273 is observed to drop down at $\nu_{\mathrm{rest}} \sim 10^{15.6}\,\mathrm{Hz}$ and increases again at $\sim 75\,\mathrm{Ryd}$ (Kriss et al. \cite{krissetal}).
The behaviour of the SED of HE~2347-4342 in this energy range is totally ignored, since it is not known as X-ray source.

\subsection{Results}
\label{sec_results}

   \begin{figure}
   \centering
   \resizebox{\hsize}{!}{\includegraphics[bb=160 380 560 775,clip=,angle=90]{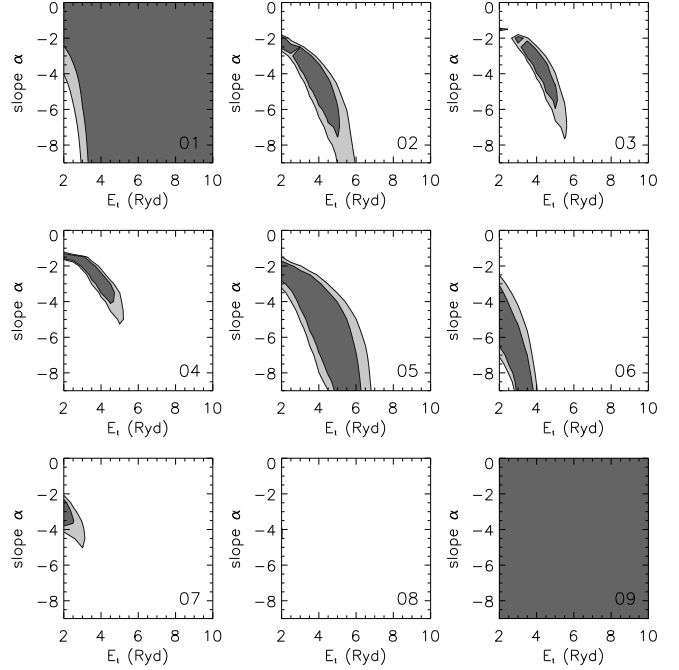}}
   \caption{Results for the parameters slope and turnover energy of the adopted SED for absorbers 1 -- 9. The numbers of the absorbers refer to the numerical order in Table \ref{clouds}. The panels show $2 \sigma$ (dark grey) and $3 \sigma$ (light grey) confidence contours, respectively.
              }
         \label{results_01-09}
   \end{figure}
   \begin{figure}
   \centering
   \resizebox{\hsize}{!}{\includegraphics[bb=295 380 560 775,clip=,angle=90]{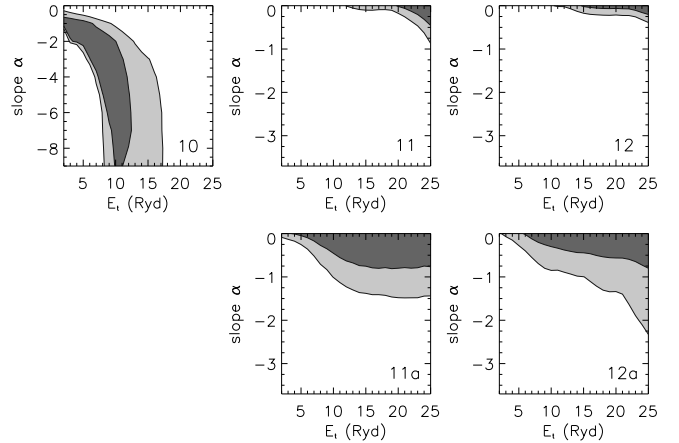}}
   \caption{$2 \sigma$  and $3 \sigma$ confidence regions of the parameters for the absorbers 10 -- 12 (upper panels). The lower panels show the same confidence regions for the high velocity absorbers 11 and 12 under the assumption of larger \ion{H}{i} column densities. Note the different axes range for absorber 10 and with respect to Fig. \ref{results_01-09}.
              }
         \label{results_10-12}
  \end{figure}
   \begin{table}
      \caption[]{Best fit results and goodness-of-fit parameters. Results for 11a and 12a were obtained under the assumption of larger \ion{H}{i} column densities. Absorbers 1 and 9 show only 2 and 3 different ions, respectively. Therefore, the $\chi^{2}$ is low for all parameters and a reliable description is impossible.}
         \label{results_table}
     $$ 
         \begin{array}{c l r r r }
            \hline
            \noalign{\smallskip}
             \mathrm{Absorber} & Q(\nu|\chi^{2}) & E_{\mathrm{t}} (\mathrm{Ryd}) & \alpha & \log U \\
            \noalign{\smallskip}
            \hline
            \noalign{\smallskip}
 1   &  1.0           &    ?  &   ?  &   ?   \\
 2   &  0.9302        &   3.0 & -3.0 &  0.10 \\
 3   &  0.9098        &   4.0 & -3.0 & -0.85 \\
 4   &  0.9835        &   4.0 & -3.0 & -1.10 \\
 5   &  0.9852        &   4.0 & -4.0 & -0.15 \\
 6   &  0.9900        &   3.0 & -8.0 &  0.50 \\
 7   &  3 \cdot 10^{-6}  &   3.0 & -4.0 & -0.55 \\
 8   &  0.0           &   3.0 & -7.0 &  0.20 \\
 9   &  1.0           &    ?  &   ?  &   ?   \\
 10  &  0.4699        &   8.0 & -3.0 &  0.05 \\
 11  &  1 \cdot 10^{-17} &  25.0 &  0.0 & -2.10 \\
 12  &  2 \cdot 10^{-11} &  25.0 &  0.0 & -2.15 \\
 \noalign{\smallskip}                
 11\mathrm{a} &  0.0016        &  25.0 &  0.0 & -1.25 \\
 12\mathrm{a} &  0.0033        &  25.0 &  0.0 & -2.20 \\
            \noalign{\smallskip}
            \hline
         \end{array}
     $$ 
   \end{table}


The results are illustrated in Figs. \ref{results_01-09} and \ref{results_10-12}.
The contours of 95.4\,\% ($2 \sigma$) and 99.73\,\% ($3 \sigma$) confidence are shown.
Since the incident continuum shapes are only determined by the position of the turnover and the steepness of the slope, different combinations of these two parameters with the same ionizing flux can lead to similar column densities. 
The parameters of the SED resulting from the best fits of each absorber and the corresponding values for the goodness-of-fit are listed in Tab. \ref{results_table}.
We discuss the absorbers following the observed grouping in the spectrum.

Constraints derived from absorber 1 are very weak, since only two column density ratios contribute to the fit, each of them with large error bars.
Consequently, this absorber is excluded from further discussion. 
Similar regions with turnover energies less than 6 Ryd are covered by absorbers 2 -- 5 with $2 \sigma$ confidence. 
Altogether, absorbers 1 -- 5 constrain the parameters of the incident spectrum to have a turnover energy of $\sim 4.0$\, Ryd with a slope of about $-3$.
The inclusion of the \ion{Si}{iv} column densities reveals that the modelled values for absorbers 3 and 4 exceed the observed upper limits by about 10\,\% and flatter slopes would be preferred.

For absorber 6 the $2 \sigma$ confidence region is located at the lower left part of the parameter space with respect to absorbers 2--5.
Despite the higher velocity it may belong to the group 1 -- 5 even though the minization of the $\chi^{2}$-function leads to $\alpha \sim -8$, a significantly steeper slope.
For these parameters the modelled \ion{N}{v} column density exceeds the observed upper limit by $\sim 13$\,\%.
Considering the spectral parameters of absorber 10, it might even belong to this complex of absorbers but with an about 3 times larger velocity.
Its confidence contours are shown in the left panel of Fig. \ref{results_10-12}. The $2 \sigma$ region reaches from flat slopes with low turnover energies to steep spectra with $E_{\mathrm{t}} \sim 10\,\mathrm{Ryd}$.
We yield the best fit at $\alpha \sim -3$ and a turnover energy of 8 Ryd with 47\,\% confidence.   

The results derived from absorbers 2--6 lead to $Q > 0.90$ in all cases implying high statistical significance.
We find larger values of $Q$ at decreasing minimal $\chi^{2}$ and less degrees of freedom.
This behaviour would be exspected since less measured column density ratios lead to better fits although not necessarily to better models if the observed limits are included.
Thus, the derived goodness-of-fit parameters become more meaningful the larger the number of contributing ions are, or in connection with the LC check results.  

The constraints of the radiation field parameters derived from absorbers 7 -- 9 are inconsistent considering either the fit results or the upper limit restrictions.
Number 7 demands low turnover energies and a slope of about $-3$ with $3 \sigma$ confidence.
The LC check restricts the allowed parameters region to slopes flatter than $-2.0$ at the same tunover energies.
The analysis of absorber 8 leads to $\alpha \sim -4$ and $E_{\mathrm{t}} \sim 2\,\mathrm{Ryd}$, where the modelled \ion{Si}{iv} column density is about 9\,\% larger than the observed limit.
The value of $Q$ is in the order of $10^{-20}$ indicating problems with the adopted model.
Since the SED is observed to increase up to 3 Ryd, the estimated turnover energy is unrealistic.
Considering this, the best model rises up to $\sim 3\,\mathrm{Ryd}$ but yields no countable goodness-of-fit parameter. 

No confidence contours are plotted for absorber 9 since only two measured ions lead to sufficient small $\chi^{2}$-values for all parameters. 
The LC check considering \ion{Si}{iv} points to the upper right part of our parameter region, where the modelled column densities of \ion{C}{iv} and \ion{N}{v} exceed the observed limits up to about 12\,\%.
Their LC checks lead to models with $\alpha = 0$.
Since only two measured column density values mean one ratio constituting the fit, the obtained results are missing any statistical significance and will not be further discussed.

Considering absorbers 7 and 8, the ionizing radiation source seems to turn over at $\sim 2\,\mathrm{Ryd}$\, with a slope about $-3 \dots -4$.
This contradicts the observed flux distribution (Fig. \ref{sed}).
Even the best results from the fitting procedure suffer from high $\chi^{2}$-values and lead to large residuals, indicating problems with the model asumptions. 
Applying the LC check best results would be expected at high turnover energies or flat spectral indices, contradicting the fitting results.
Especially the estimated column densities of system 8 are supposed to be very accurate since all ions show unblended absorption features as can be seen from Fig. \ref{asv}.
Thus, the effective ionizing radiation might be different from the model.
Possibly, the QSO radiation is strongly shielded by the absorbers 9 -- 12.
If collisional ionization or shock heating were important completely different results would be expected.

   \begin{figure}
   \centering
   \includegraphics[bb=430 640 560 775,clip=,width=4.5cm,angle=90]{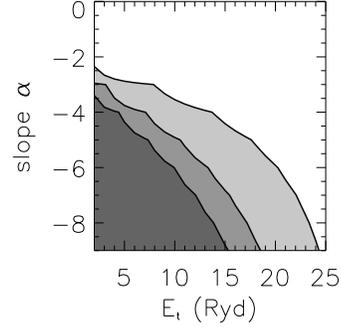}
   \caption{Parameter combinations producing a flux below RASS detection limit $2 \cdot 10^{-13}\,\mathrm{erg\,s}^{-1}\,\mathrm{cm}^{-2}$, below $10^{-12}$ and $10^{-11}\,\mathrm{erg\,s}^{-1}\,\mathrm{cm}^{-2}$.
              }
         \label{RASSflux}
  \end{figure}
For the high velocity lines, absorbers 11 and 12, we find flat slopes and large turnover energies shown in the upper right panels of Fig. \ref{results_10-12}.
The spectral flatness of the most confident continua at the highest turnover energies considered in our model grid and the LC check indicate that even 25 Ryd does not describe the SED correctly and spectra with a turnover at higher energies would improve the fit.
This idea is supported as well by the very low confidence of the best fits, yielding goodness-of-fit parameters of $10^{-17}$ and $10^{-11}$, respectively.
The energy of 25 Ryd corresponds to a wavelength of $\sim 37\,\mathrm{\AA}$, which is already in the spectral range of soft X-rays.
The ROSAT all-sky survey (RASS) detected no X-ray source at the relevant position within an error circle less than $1 \deg$.
RASS detected sources in the energy band 0.1--2.4 kev corresponding to a wavelength range of 5--124 \AA.
In the QSO rest frame this corresponds to 1.3--32.0 \AA\, or 28.5--682 Ryd, respectively.
Thus, the modelled energy of turnover is slightly lower than the energy range observed with ROSAT.
In our case the detection limit of RASS is about $2 \cdot 10^{-13}\,\mathrm{erg\,cm}^{-2}\,\mathrm{s}^{-1}$ for the integrated flux.
Adopting $q_{0} = 0.5,\, H_{0} = 75\,\mathrm{km\,s}^{-1}\mathrm{Mpc}^{-1}$, $\alpha$ has to be steeper than $-30$ to produce a flux low enough to be not detected by ROSAT.
Figure \ref{RASSflux} shows the parameter combinations consistent with the non-detection in RASS.
Indeed, none of our confident models describes the observations consistently.
The reason might be a variability of the QSO's X-ray radiation, which is not unusual in high luminosity quasars. 
Therefore, the X-ray flux might have been higher at the time when the optical observations were made. 

Another reason to reject models with flat slopes and high turnover energies is the non-detection of various ions in the UV data.
For instance the model with a turnover at 25 Ryd and $\alpha = 0$ predicts strong \ion{Mg}{viii} absorption comparable to the column density of \ion{C}{iv} which is not identified in the FOS data. 

The derived hydrogen column densities of the absorbers 11 and 12 are uncertain since even Ly$\epsilon$ is still saturated in both cases.
To estimate the order of the error we compute models for the Lyman series up to \ion{H}{i} $\lambda\,921$ with different column densities.
Values exceeding $\log N_{\ion{H}{i}} \sim 17.0$, which is 1 dex larger than the estimated column densities, can be excluded since we definitely observe no Lyman limit.
We find that values 0.1 dex larger are still consistent with the data.   
A new run of the fit procedure adopting values for the column densities increased by 0.1 dex yields confidence levels as shown in the lower panels of Fig. \ref{results_10-12} named as absorbers 11a and 12a.
Compared to the first fit, we find confident models with lower turnover energies and steeper slopes.
However, best models are still those with flat slopes, yielding now significant higher but still low confidences with $Q = 0.002$ and $0.003$, respectively.

As mentioned in Sect. \ref{remarks} the \ion{O}{vi} column densities adopted for absorbers 11 and 12 are very unreliable.
For this reason they have not been included in the fit.
Considering the modelled \ion{O}{vi} column densities in the $3 \sigma$ regions, we find values in the range of $\log N_{\mathrm{\ion{O}{vi}}} \sim 13.5 \dots 14.0$ for absorbers 11, 12, and 12a, while 11a yields $15.0 \dots 15.5$ due to a different ionization parameter.
All these values are consistent with the observations.

Since the high velocity absorbers 11 and 12 show the strongest lines in \ion{H}{i} and possibly \ion{O}{vi} and seem to be exposed to a harder continuum than the other absorbers they are likely shielding the others from QSO radiation.
To be responsible for the absence of the proximity effect absorbers 11 and 12 are supposed to produce strong \ion{He}{ii} lines which are optically thick in the \ion{He}{ii} continuum.
For our most confident models we find helium column densites of $\log N_{\mathrm{\ion{He}{ii}}} \sim 16.3$ and a slightly larger value in case of absorber 12a.
Models producing the expected column density of $\log N_{\mathrm{{He}{ii}}} \ge 17.0$ lead to turnover energies less than 4 Ryd covering a parameter region consistent with the ROSAT non-detection but outside the $3 \sigma$ confidence contours.
However, none of the absorbers 1 -- 10 can be desribed consistently with models producing enough \ion{He}{ii} to be responsible for a shielding effect.

\subsection{Comparison to Mathews--Ferland continuum}

   \begin{figure}
   \centering
   \resizebox{\hsize}{!}{\includegraphics[bb=25 320 550 655,clip=]{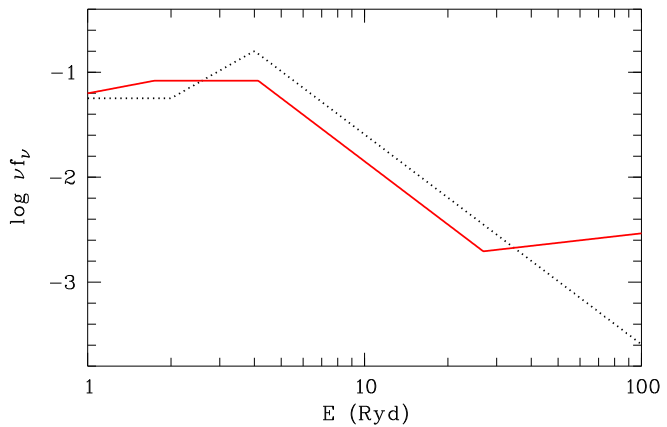}}
   \caption{AGN type spectrum from Mathews \& Ferland (\cite{mathewsferland}; solid line) in comparison to a model with $\alpha = -3.0$ and a turnover energy of 4.0 Ryd (dotted). 
              }
         \label{mf_sed}
   \end{figure}

Mathews \& Ferland (\cite{mathewsferland}; hereafter MF) deduced an AGN type energy distribution from observations and photoionization models of AGN broad emission lines.
A characteristic feature of their spectrum is a drop down at $4.13\,\mathrm{Ryd}$ with an $\alpha = -3.0$ slope (Fig. \ref{mf_sed}).
For comparison we have also considered the MF continuum as ionizing source.

Absorbers 1 -- 5 can be fitted consistently with a MF continuum.
The modelled \ion{Si}{iv} column density of numbers 1, 3 and 4 exceeds the observed upper limits by 1\,\%, 6\,\%, or 10\,\%, respectively.
Since the the MF continuum is similar to our model with $E_{\mathrm{t}} = 4.0\, \mathrm{Ryd}$ and $\alpha = -3.0$ equivalent results should be expected.
The ionization parameters required by the Mathews--Ferland model differ from our model by less than $\Delta \log U = 0.15$. 

Absorbers 6 -- 9 can hardly be described using the MF continuum.
The best fits lead to a metal deficiency of about 1 dex compared to the observations.
To improve the fit, we have varied the metalicity in a range of 1 to 10 times solar and find that the higher metalicities lead to better fits.
An exception is silicon since our models overestimate the column density of \ion{Si}{iv} of the absorbers 7 and 8 by 20\,\% and 8\,\%, respectively.
Column densities of \ion{N}{v} show the tendency towards larger values than observed indicating lower nitrogen abundances of the absorbers 6 -- 8.
Another possibilty to improve the fit quality could be an increase of the carbon abundance, since the deficiency is especially strong for \ion{C}{iii} and \ion{C}{iv}.
These results suggest to vary the element abundances to improve the power law models.
Due to the lack of different elements and ionization stages this would only increase the number of free parameters and lead to additional uncertainties.
Absorber 9 suffers from the shortage of observed elements and cannot be modelled consistently with the MF continuum. 
  
Best fits of the absorbers 11 and 12 to the MF continuum suffer from large $\chi^{2}$-values and $Q \sim 0.0$ for both absorbers indicating problems with the input model.
If the absorbers 11 and 12 are photoionized by the quasar as dominating source, the SED of HE~2347-4342 is different from the MF continuum.
On the other hand a description of absorber 10 is possible without enhanced metalicity supporting the idea that absorber 10 is physically connected to the complex of absorbers 1 -- 5, even though the best fit yields $Q \sim 10^{-9}$.

\subsection{Collisional ionization models}

   \begin{figure}
   \centering
   \resizebox{\hsize}{!}{\includegraphics[bb=40 415 355 670,clip=]{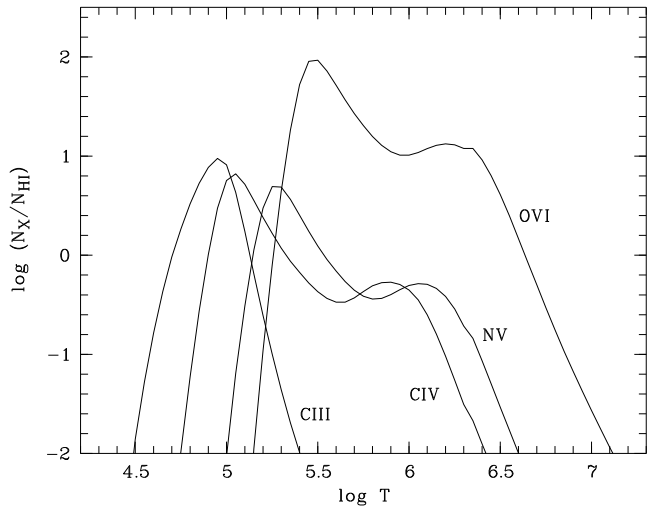}}
   \caption{Density ratios with respect to \ion{H}{i} as function of temperature $T$ for a highly ionizied gas in collisional equilibrium. The ionization data were taken from Sutherland \& Dopita (\cite{sutherlanddopita}) and scaled to solar abundances in accordance with Grevesse \& Sauval (\cite{grevessesauval}).
              }
         \label{collisional}
   \end{figure}

The adopted photoionization models failed to describe consistently the observed column density ratios of absorbers 8, 11, and 12.
In a scenario where the gas was ejected by the central source and is now falling back again, collisional ionization may play a dominant role.
Figure \ref{collisional} shows the density ratios relative to \ion{H}{i} of all observed ions, based on the ionization fractions for collisional ionization equilibrium from Sutherland \& Dopita (\cite{sutherlanddopita}).

However, the measured column densities cannot be explained with a single temperature model.
The best fit for absorber 8 is obtained at $\log T = 5.30$, though carbon and oxygen are significantly overestimated.
The main constraining ion in this case is \ion{O}{vi}, whose column density fraction increases steeply with temperature up to $300\,000\,\mathrm{K}$ (see Fig. \ref{collisional})
Considering all ions separately, temperatures in the range of $\log T \sim 5.05$ (\ion{C}{iv}) to 5.35 (\ion{O}{vi}) are found.
Since a single temperature model likely oversimplifies the real physical conditions, a mixture of temperatures of $100\,000 < T < 300\,000\,\mathrm{K}$ may explain the observed column density ratios.
However, fitting a multiphase model to the observations is not possible unambiguously for the presented data, although absorber 8 provides the most reliable column density measurements.

Describing absorbers 11 and 12 with collisionally ionized gas is not possible either.
The observed column density ratios are located on the steeper parts of the functions shown in Fig. \ref{collisional}.
Thus, the resulting temperatures are spread over a range of about an order of magnitude in the lower temperature part ($\log T \sim 4 \dots 5$) as well as at higher temperatures ($\log T \sim 6 \dots 7$).
Though the observations of absorbers 11 and 12 might be explained by a mixture of photo and collisionally ionized gas, a realistic multiphase model would introduce additional free parameters and is beyond the scope of the present study.
A further possibility is that ionization is not in equilibrium.

\section{Conclusions}

We have analyzed the complex associated system of HE~2347-4342 showing absorption features of \ion{H}{i}, \ion{C}{iii}, \ion{C}{iv}, \ion{N}{v}, and \ion{O}{vi} in the optical.
The state of line locking apparently visible in \ion{C}{iv} and \ion{N}{v} was investigated.
Comparing simulated spectra with randomly distributed absorption doublets we found no statistical evidence for the presence of this effect.
We point out that apparently present line locking has to be confirmed quantitatively.

We have used column densities of visible ions and upper limits of non-detected ions redwards from the QSO emission line redshift to model its SED adopting a simple model matching the observed low energy part.
Applying our method to the data leads to widely consistent results.
Five absorbers within the velocity range $200 < v < 600\, \mathrm{km\,s}^{-1}$ can be modelled with a confidence of at least 90\,\% with a spectral slope $\alpha \sim -3$ at the energy $\sim 3 \dots4$\,Ryd resulting in a SED similar to a Mathews--Ferland continuum.
Two other absorbers at 850 and $1259\, \mathrm{km\,s}^{-1}$ satisfy similar models with a steeper slope (99\,\% confidence, absorber 6) or a higher turnover energy (47\,\% confidence, absorber 10), respectively.
Thus, the velocities of the absorbers and their distance to the quasar seem to show no correlation.

We failed to find parameter combinations to describe the absorption features in the range $900 < v < 1200\, \mathrm{km\,s}^{-1}$.
We conclude that these absorbers are either not purely photoionized or the effective ionizing continuum is not compatible with our model assumptions.
Computations with the MF continuum indicate a higher metalicity than solar with reduced silicon abundance. 
Differential abundance effects are not considered within our simple model, since additional parameters would cause a higher degree of freedom and increase the uncertainties.

Collisional ionization at a single gas temperature could not be confirmed as dominating process for these absorbers.
We found that the absorber at $1033\,\mathrm{km\,s}^{-1}$ may be described with a mixture of temperatures of $100\,000 < T < 300\,000\,\mathrm{K}$.
However, the presented data are not sufficient to constrain multiphase gas models.

A further complication is the possible variability of the UV continua of luminous quasars in strength and shape.
It has recently been discovered that the UV continuum of the similar luminous QSO HS~1700+6416 ($z = 2.73$) has varied by a factor of 3 to 4 at $1200\,\mathrm{\AA}$, while in the optical the amplitude is only $0.1^{\mathrm{m}}$ (Reimers et al., in preparation).
Consequently, the shape of the observed UV continuum (Fig. \ref{sed}) might be different from the one responsible for photoionization of the associated system.
However, new observations made in June 2003 with the UVES spectrograph with the same resolution but an improved signal-to-noise ratio reveal no variability of the observed components of \ion{O}{vi} and \ion{N}{v}.
The time interval between the observations analyzed in this paper and the new data is about 2.5 yr, corresponding to about 8 months in the QSO rest frame.

Modelling the two high velocity absorbers ($1300 < v < 1600\,\mathrm{km\,s}^{-1}$) we found best results with flat slopes at high turnover energies.
However, an energy distribution like this is in conflict with the non-detection of HE~2347-4342 in the ROSAT all sky survey as well as the non-detection of \ion{Mg}{viii} in the FOS portion of the spectrum.
Assuming the high velocity absorbers are the closest to the quasar they are expected to be illuminated by a hard UV continuum.
We confirm a spectral energy distribution harder than the MF continuum.
Since the two high velocity absorbers lead to the hardest continuum and large \ion{He}{ii} column densities ($\log N_{\mathrm{\ion{He}{ii}}} > 16.3$), they may shield the other absorbers from the QSO radiation.
If they are optically thick in the \ion{He}{II} continuum, lower velocity absorbers are exposed to a filtered, softer radiation field, as observed.

The results could be improved if more ions were included in the analysis.
Therefore, column densities of ions observed in the UV portion of the spectrum like \ion{O}{v} have to be estimated more accurately.
In the available FOS and GHRS spectra the resolution and $S/N$ do not allow to determine column densities for individual absorbers.
Especially, the models for the high velocity clouds which suffer from saturated hydrogen and uncertain \ion{O}{vi} absorption features could be improved, if future UV observations allow to resolve further lines like \ion{O}{v} 630\,\AA, \ion{O}{iv} 608\,$\mathrm{\AA}$ etc.

In order to constrain the high energy portion of the SED more precisely, several ionizations stages of neon (\ion{Ne}{v} -- \ion{Ne}{viii}) could be used, which are expected to be visible in the UV spectral range.
If these lines are observed with a sufficiently high resolution and signal-to-noise ratio, e.g. by the Cosmic Origin Spectrograph (COS) on HST, the energy range covered by the ionization potentials of the observed ions will be extended by a factor of 2.
Thus, the high energy portion of the SED could be modelled more confidently.

\begin{acknowledgements}
We wish to thank the anonymous referee who helped to improve the paper.
This work has been supported by the Verbundforschung (DLR) of the BMBF under Grant No. 50 OR 0203.
\end{acknowledgements}


\begin{thebibliography}{}

   \bibitem[2000]{ballester} Ballester, P., Boitquin, O., et al. 2000,
      ESO Mess., 101, 31

   \bibitem[1989]{braunmilgrom} Braun, E. \& Milgrom, M. 1989,
      ApJ, 342, 100

   \bibitem[1997]{cloudy} Ferland, G. 1997,
      A Brief Introduction to Cloudy 
      (Internal Rep., Lexington: Univ. Kentucky) 

   \bibitem[1986]{foltz86} Foltz, C. B., Weymann, R. J., Peterson, B. M., 
     Sun, L., Malkan, M. A. \& Chaffee, F. H. 1986,
      ApJ, 307, 504

   \bibitem[1987]{foltzetal} Foltz, C. B., Weymann, R. J., Morris, S. L. 
     \& Turnshek, D. A. 1987,
      ApJ, 317, 450

   \bibitem[1988]{foltz88} Foltz, C. B., Chaffee, F. H., Weymann, R. J. 
     \& Anderson, S. F. 1988,
      Proceedings of the QSO Absorption Line Meeting, 53

   \bibitem[1998]{grevessesauval} Grevesse, N. \& Sauval, A. J. 1998,
      Space Sci. Rev., 85, 161

   \bibitem[1993]{hamannferland93} Hamann, F. \& Ferland, G. 1993,
      ApJ, 418, 11

   \bibitem[1999]{hamannferland} Hamann, F. \& Ferland, G. 1999,
      ARA\&A, 37, 487

   \bibitem[1999]{krissetal} Kriss, G. A., Davidsen, A. F., Zheng, W. 
     \& Lee, G. 1999,
      ApJ, 527, 683

   \bibitem[2001]{krissetal2001} Kriss, G. A., Shull, J. M., Oegerle, W., 
     et al. 2001,
     Science, 293, 1112

   \bibitem[1987]{mathewsferland} Mathews, W. G. \& Ferland, G. J. 1987,
      ApJ, 323, 456

   \bibitem[1997]{reimerswisotzki} Reimers, D. \& Wisotzki, L. 1997,
      The Messenger, 88, 14

   \bibitem[1997]{reimers97} Reimers, D., K\"ohler, S., Wisotzki, L., 
     Groote, D., Rodriguez-Pascual, P. \& Wamsteker, W. 1997,
      A\&A, 327, 890

   \bibitem[1998]{reimers98} Reimers, D., K\"ohler, S., Hagen, H.-J. \& 
     Wisotzki, L. 1998, 
      in Ultraviolet Astrophysics Beyond the \textit{IUE} Final Archive, 
      ed. W. Wamsteker \& R. Gonzalez Riestra (Noordwijk, ESA), 579

   \bibitem[1973]{scargle} Scargle, J. D. 1973,
      ApJ, 179, 705

   \bibitem[2002]{smette} Smette, A., Heap, S. R., Williger, G. M., Tripp, 
     T. M., Jenkins, E. B. \& Songaila, A. 2002,
      ApJ, 564, 542
    
   \bibitem[2000]{srianand} Srianand, R. 2000,
      ApJ, 528, 617

   \bibitem[2002]{srianandetal} Srianand, R., Petitjean, P., Ledoux, C. 
     \& Hazard, C. 2002,
      MNRAS, 336, 753

   \bibitem[1993]{sutherlanddopita} Sutherland, R. S. \& Dopita, M. A. 1993,
      ApJS, 88, 253

   \bibitem[2002]{telfer} Telfer, R. C., Zheng, W., Kriss, G. A. \&
     Davidsen, A. F. 2002,
      ApJ, 565, 773

   \bibitem[2001]{vilkoviskij} Vilkoviskij, E. Y. \& Irwin, M. J. 2001,
      MNRAS, 321, 4

   \bibitem[1993]{wampler} Wampler, E. J., Bergeron, J. \& Petitjean, P. 1993,
      A\&A, 273, 15

   \bibitem[1979]{weymann} Weymann, R. J., Williams, R. E., Peterson, B. M. \&
     Turnshek, D. A. 1979,
      ApJ, 234, 33

   \bibitem[1997]{zheng} Zheng, W., Kriss, G. A., Telfer, R. C., Grimes, J. P.
     \& Davidsen, A. F. 1997, 
      ApJ, 475, 469 

\end{thebibliography}
\end{document}